\documentclass[reprint,noshowpacs,noshowkeys,prd,balancelastpage,nofootinbib]{revtex4}
\usepackage[utf8]{inputenc}
\usepackage{color}
\usepackage{amsmath}
\usepackage{amssymb}
\usepackage{graphicx}
\usepackage[unicode=true,
 bookmarks=false,
 breaklinks=false,pdfborder={0 0 1},backref=section,colorlinks=true]
 {hyperref}
\hypersetup{
 linkcolor=blue,urlcolor=blue,citecolor=red}

\makeatletter
\@ifundefined{textcolor}{}
{%
 \definecolor{BLACK}{gray}{0}
 \definecolor{WHITE}{gray}{1}
 \definecolor{RED}{rgb}{1,0,0}
 \definecolor{GREEN}{rgb}{0,1,0}
 \definecolor{BLUE}{rgb}{0,0,1}
 \definecolor{CYAN}{cmyk}{1,0,0,0}
 \definecolor{MAGENTA}{cmyk}{0,1,0,0}
 \definecolor{YELLOW}{cmyk}{0,0,1,0}
}

\usepackage{color}
\usepackage{amsfonts}
\usepackage{mdframed}
\usepackage{footnote}
\usepackage{graphicx}
\usepackage{float}
\usepackage[font=footnotesize,it]{caption}
\usepackage{tikz}
\usepackage{tikz-3dplot}

\@ifundefined{textcolor}{}{
 \definecolor{BLACK}{gray}{0}
 \definecolor{WHITE}{gray}{1}
 \definecolor{RED}{rgb}{1,0,0}
 \definecolor{GREEN}{rgb}{0,1,0}
 \definecolor{BLUE}{rgb}{0,0,1}
 \definecolor{CYAN}{cmyk}{1,0,0,0}
 \definecolor{MAGENTA}{cmyk}{0,1,0,0}
 \definecolor{YELLOW}{cmyk}{0,0,1,0}
}

\makeatother

\begin{document}
\title{A Geometrical Model for the Evolution of Spherical Planetary Nebulae
Based on Thin-Shell Formalism}
\author{S. Danial Forghani}
\email{danial.forghani@final.edu.tr}

\affiliation{Faculty of Engineering, Final International University, Kyrenia, North
Cyprus via Mersin 10, Turkey}
\author{Ibrahim Gullu}
\email{ibrahim.gullu@emu.edu.tr}

\affiliation{Department of Physics, Faculty of Arts and Sciences, Eastern Mediterranean
University, Famagusta, North Cyprus via Mersin 10, Turkey}
\author{S. Habib Mazharimousavi}
\email{habib.mazhari@emu.edu.tr}

\affiliation{Department of Physics, Faculty of Arts and Sciences, Eastern Mediterranean
University, Famagusta, North Cyprus via Mersin 10, Turkey}
\date{\today }
\begin{abstract}
A spherical planetary nebula is described as a geometric model. The
nebula itself is considered as a thin-shell which visualized as a
boundary of two spacetimes. The inner and outer curvature tensors
of the thin-shell are found in order to get an expression of the energy-momentum
tensor on the thin-shell. The energy density and pressure expressions
are derived using the energy-momentum tensor. The time evolution of
the radius of the thin-shell is obtained in terms of the energy density.
The model is tested by using a simple power function for decreasing
energy density and the evolution pattern of the planetary nebula is
attained.
\end{abstract}
\maketitle

\section{Introduction}

Planetary nebulae (PNe; singular, PN) are one of the last stages in
the low and intermediate-mass stellar evolution which provide a transition
between two drastically different phases of a dying star; a red giant
and a white dwarf. A typical planetary nebula, consists of an ever
expanding low-density ionized gas and a central hot star, known as
the PN nucleus (PNN; a.k.a. PN central start, PNCS). The typical density
of a PN is about 100 to 10,000 particles per cubic centimeter \cite{Osterbrock1}.
Widely different in shapes and features, morphologies of PNe are heavily
influenced by the characteristics of the post-PN phases, the PNN,
and the environmental factors. Although the physics behind the dynamical
processes leading to this vast morphological variety is not well-understood
yet, some factors such as rotation \cite{Chita1,Garcia2}, mass density
distribution \cite{Frank3}, rotation \cite{Garcia1}, metalicity
and magneticity of the proginator asymptotic giant branch (AGB) star
\cite{Garcia1,Soker1,Kingsburgh1,Frank2,Garcia3,Matt1}, the stellar
winds by the PNN itself and the neighboring stars \cite{Frank2,Kwok1,Balick1,Icke1},
and a possible binary-system occurrence and their corresponding tidal
interactions \cite{Morris1,Morris2,Soker2,Nordhaus1,Tocknell1,Jones1}
are enumerated in the literature \cite{Balick2}. With a life-span
of a few tens of thousands of years, the PN-phase of a star is considerably
short compared to its billion-year overall evolution period. Yet,
its significance in our underacting of the evolution of the stars
is singular for many reasons. Firstly, over $90\%$ of stellar evolutions
somehow experience a PN phase \cite{Zijlstra1}. Secondly, PN mechanism
is one through which the chemical abundance of the interstellar medium
is evolved \cite{Kingsburgh1,Kwitter1}. Thirdly, their frequent occurrence
and versatile shapes allow us to analyze them in large groups in order
to develop our theoretical and experimental understanding of the way
of the hydrodynamics of the stellar evolution. Furthermore, extragalactic
PNe offer ways to derive stellar formation rates and metalicity gradients
in galaxies \cite{Bertolami1,Coccato1,Magrini1,Buzzoni1}. Finally,
the PN luminosity function \cite{Ciardullo1,Ciardullo2,Ciardullo3}
and the $S_{H_{\alpha}}-r$ relation \cite{Frew1}, can serve as an
accurate distance indicators up to $20\,Mpc$.

With a visual classification \cite{Manchado1}, around $20\%$ of
all PNe are round \cite{Pradhan1}. However, one should beware of
projection effects \cite{Frank2} which may cause taking a PN of a
different morphology (elliptical or bipolar) for a spherical or round,
when observed from a wrong angle. Considering the projection effects,
it is estimated that $10-20\%$ of PNe are nearly or absolutely spherical
\cite{Soker1}. These type of PNe occur more frequently among low
mass stars ($\leq1.1M_{\odot}$) \cite{Zuckerman1,Corradi1}. Although
spherical PNe expand almost uniformly and homogeneously, they possess
microstructures, in general \cite{Frank3}. However, there exists
a rare type of spherical PN which does not exhibit a microstructure,
whatsoever. The best example of this type is Abell 39, a simple spherical
shell of ionizing gas with a low brightness in the constellation of
Hercules \cite{Abell1}. In the present study, we do not intend to
study the evolution of a spherical PN by the hydrodynamical processes
it goes through. Instead, our intention is to present a simple analytical
model in which the time evolution of the radius, energy density and
pressure of such PNe can be derived by a relativistic geometrical
method. In our model, the thin outer rim of the bubble of a PN of
type Abell 39 is illustrated by a thin-shell, in its general relativistic
sense. This thin-shell, which is essentially a three-dimensional hypersurface,
divides the spacetime into two inherently different manifolds; the
interior and the exterior. For this particular study, we consider
an uncharged Vaidya's metric \cite{Vaidya}, characterized by a radial
flow of electrically neutral unpolarized matter radiation as the interior,
and a Schwarzschild metric corresponding to a gravitational field
caused by an uncharged, non-rotating spherically symmetric mass as
the exterior. The two spacetimes are connected via proper junction
conditions \cite{Israel} which certify the energy density and angular
pressure of the matter distributed over the .

The layout of the paper is as follows: In Section-II, the spacetimes
separated by the thin-shell are given and by use of Darmois-Israel
junction conditions the metric tensor on the surface of thin-shell
is defined. Moreover, the inner and outer curvature tensors are calculated
in order to get the energy-momentum tensor on the shell. By means
of the energy-momentum tensor the energy density and the pressure
of the thin-shell are calculated using the perfect fluid assumption.
Furthermore, the time evolution of the radius of the spherical thin-shell
is given at the end of this section. The time evolution of the radius
of the thin-shell is investigated assuming a simple power function
in Section-III. The paper is brought to completion with a conclusion
in Secion-IV. 

\section{The Spacetimes}

A thin-shell can be visualized as a boundary separating two spacetimes,
namely an interior spacetime from an exterior. For the thin-shell
to be physical, certain junction conditions must be satisfied at the
location of the thin-shell. Only it is under these conditions that
one can claim the whole spacetime (including the inner and outer spacetimes
plus the thin-shell itself) is a solution of Einstein field equations.
These conditions, which are known as Darmois-Israel junction conditions
in general relativity, firstly, demand the metric tensor to be continuous
across the thin-shell. In our consideration, this thin-shell will
be corresponding to a spherically symmetric PN with the defining equation
\begin{equation}
\Sigma:=r_{\pm}-a\left(\tau\right)=0,\label{defining_equation}
\end{equation}
and intrinsic metric

\begin{equation}
ds_{\Sigma}^{2}=g_{ij}d\xi^{i}d\xi^{j}=-d\tau^{2}+a^{2}\left(\tau\right)\left(d\theta^{2}+\sin^{2}\theta\,d\phi^{2}\right),\label{intrinsic_metric}
\end{equation}
where $\xi^{i}=\left(\tau,\theta,\phi\right)$ are the coordinates
on the thin-shell and $a\left(\tau\right)$ stands for the radius
of the hypersurface $\Sigma$ dividing the spacetime into two distinct
$3+1$-dimensional manifolds of class $C^{4}$. Also, $r_{\pm}$ are
the radial coordinates of the outer $\left(+\right)$ and the inner
$\left(-\right)$ spacetimes. The first junction condition requires
the satisfaction of 
\begin{equation}
\left(ds_{-}^{2}\right)_{\Sigma}=\left(ds_{+}^{2}\right)_{\Sigma}=ds_{\Sigma}^{2}
\end{equation}
on the boundary. In this study, we model the interior spacetime by
the Vaidya metric 
\begin{equation}
ds_{-}^{2}=-f_{-}\left(r_{-},t_{-}\right)dt_{-}^{2}+2\epsilon\,dt_{-}\,dr_{-}+r_{-}^{2}\left(d\theta^{2}+\sin^{2}\theta\,d\phi^{2}\right),\label{Vaidya_metric}
\end{equation}
where 
\begin{equation}
f\left(r_{-},t_{-}\right)=1-\frac{2M_{-}\left(t_{-}\right)}{r_{-}},\label{Vaidya_function}
\end{equation}
is the interior metric function, in which $M_{-}\left(t_{-}\right)$
is the time-dependent mass of the central gravitational object. Furthermore,
$\epsilon$ takes on $-1$ and $+1$ for outgoing and incoming waves,
respectively. The outer spacetime, on the other hand, will be of Schwarzschild-type
geometry, given with the line element 
\begin{equation}
ds_{+}^{2}=-f_{+}\left(r_{+},t_{+}\right)dt_{+}^{2}+f_{+}^{-1}\left(r_{+},t_{+}\right)dr_{+}^{2}+\rho_{+}^{2}\left(d\theta^{2}+\sin^{2}\theta\,d\phi^{2}\right),\label{Schwarzschild metric}
\end{equation}
where the metric function is 
\begin{equation}
f_{+}\left(r_{+},t_{+}\right)=1-\frac{2M_{+}}{r_{+}},\label{Schwarzschild function}
\end{equation}
with constant mass $M_{+}$. Note that as a direct consequence of
the first junction condition, we have $\theta_{-}=\theta_{+}=\theta$
and $\phi_{-}=\phi_{+}=\phi$ at the shell. As another important consideration,
note that the radius of the shell $a\left(\tau\right)$ must be chosen
such that it exceeds the event horizons of both inner and outer spacetimes.
As will be seen in the following lines, we always have $M_{+}>M_{-}$,
hence, we must have $a>2M_{+}$ at all times.

To have a boundary which indeed distinguishes the inside from the
outside, the thin-shell itself must possess an energy-momentum tensor.
This tensor is given by the second junction condition, which relates
the energy-momentum tensor of the matter at the thin-shell to the
discontinuity in the second fundamental form across the shell. The
curvature tensors of the interior and exterior spacetimes are given
by \cite{Eisenhart}

\begin{equation}
K_{ij}^{\pm}=-n_{\alpha}^{\pm}\left(\frac{\partial^{2}\chi_{\pm}^{\alpha}}{\partial\xi^{i}\partial\xi^{j}}+\Gamma_{\mu\nu}^{\alpha\pm}\frac{\partial\chi_{\pm}^{\mu}}{\partial\xi^{i}}\frac{\partial\chi_{\pm}^{\nu}}{\partial\xi^{j}}\right),\label{curvature_tensors}
\end{equation}
in which $\chi_{\pm}^{\alpha}$ are the coordinates of the inner and
outer spacetimes, $n_{\alpha}^{\pm}$ are the components of the $4$-normal
to $\Sigma$ given by 
\begin{equation}
n_{\alpha}^{\pm}=\frac{\partial_{\alpha}\Sigma^{\pm}}{\left\vert g^{\beta\gamma}\partial_{\beta}\Sigma\partial_{\gamma}\Sigma\right\vert },\label{normal_vector}
\end{equation}
and $\Gamma_{\mu\nu}^{\alpha\pm}$ are the Christoffel symbols compatible
with the bulk metrics $g_{\mu\nu}^{\pm}$. Therefore, the second junction
conditions are expressed as 
\begin{equation}
-8\pi S_{i}^{j}=\left[K_{i}^{j}\right]_{-}^{+}-\delta_{i}^{j}\left[K\right]_{-}^{+},\label{second_junction_condition}
\end{equation}
where, $K=g^{\mu\nu}K_{\mu\nu}$ is the total curvature, $\delta_{i}^{j}$
is the Kronecker delta, and $\left[\,\right]_{-}^{+}$ indicates a
jump across the shell in the quantity it embraces, e.g. $\left[K_{ij}\right]_{-}^{+}=K_{ij}^{+}-K_{ij}^{-}$.
Accordingly, $S_{i}^{j}$ is the mixed energy-momentum tensor of the
matter at the shell, which for a perfect fluid picks up 
\begin{equation}
S_{i}^{j}=diag\left(-\sigma,p,p\right).\label{energy-momentum_tensor}
\end{equation}
Here, $\sigma$ is the energy density, whereas $p$ is the angular
pressure. For the purpose of this study, the perfect fluid assumption
seems reasonable since according to the Generalized Interacting Stellar
Wind model (GISW) \cite{Kwok1,Balick1,Icke1,Frank1} PNe evolve spherically
if their pole-to-equator density contrast $e\equiv\frac{\sigma_{p}}{\sigma_{E}}$,
associated with their AGB slow wind, is unity.

Inserting (\ref{energy-momentum_tensor}) and (\ref{curvature_tensors})
into (\ref{second_junction_condition}) yields the energy density
and pressure of the matter at the shell, as

\begin{equation}
\sigma=\frac{1}{4\pi a}\left(\sqrt{f_{-}+\overset{\cdot}{a}^{2}}-\sqrt{f_{+}+\overset{\cdot}{a}^{2}}\right),\label{energy_density}
\end{equation}
and 
\begin{equation}
p=\frac{1}{8\pi}\left(\frac{f_{+}^{\prime}+2\ddot{a}}{2\sqrt{f_{+}+\dot{a}^{2}}}-\frac{1}{2}\dot{t}_{-}\,f_{-}^{\prime}-\epsilon\,\frac{\ddot{t_{-}}}{\dot{t_{-}}}\right)-\frac{\sigma}{2}.\label{pressure}
\end{equation}
Herein, the overdot $\left(\dot{}\right)$ and the prime $\left(^{\prime}\right)$
stand for total derivatives with respect to the proper time $\tau$
and the radial coordinates $r_{\pm}$, respectively. Moreover, as
an auxiliary equation, one could calculate the conservation of energy
by starting from $\nabla_{j}S_{i}^{j}=0$ and setting $i\equiv\tau$.
After some calculations, the result will come out to be 
\begin{equation}
\sigma^{\prime}+\frac{2}{a}\,\left(\sigma+p\right)=\frac{1}{4\pi a}\left[\frac{\epsilon\,\overset{\cdot}{a}^{2}+\frac{1}{2}\epsilon\,f_{-}+\overset{\cdot}{a}\,\sqrt{f_{-}+\overset{\cdot}{a}^{2}}}{f_{-}^{2}\,\sqrt{f_{-}+\overset{\cdot}{a}^{2}}}\,\frac{\partial f_{-}}{\partial t_{-}}\right].\label{energy-conservation}
\end{equation}
Note that, unlike the cases in which we have static metric functions
on the two sides of the shell, here a non-zero time-dependent term
appears on the right-hand side due to the dynamic nature of the Vaidya
metric. However, if the interior central mass $M_{-}$ does not depend
on time, the right-hand side will identically amount to zero, as expected.

Besides the energy conservation relation in Eq. (\ref{energy-conservation}),
another mechanical energy-like equation can be generated by rewriting
Eq. (\ref{energy_density}) in the form 
\begin{equation}
\frac{1}{2}\overset{\cdot}{a}^{2}+V_{\text{eff}}=0,\label{mechanical_energy_equation}
\end{equation}
where $\frac{1}{2}\overset{\cdot}{a}^{2}$ resembles the kinetic term
and the effective potential 
\begin{equation}
V_{\text{eff}}=\frac{1}{2}\,\left[\frac{f_{+}+f_{-}}{2}-\left(\frac{f_{+}-f_{-}}{\kappa\sigma a}\right)^{2}-\left(\frac{\kappa\sigma a}{4}\right)^{2}\right]\label{effective_potential}
\end{equation}
is a function of the radius of the shell and the time coordinate $t_{-}$
through $f_{-}$. The gravitational mass of the exterior metric, $M_{+}$,
is the sum of the gravitational mass of the interior metric, $M_{-}$,
and the mass of the shell that is the multiplication of its energy
density by its surface area, i.e. $4\pi a^{2}\sigma$. Therefore,
we have the mass relation 
\begin{equation}
M_{-}=M_{+}-4\pi a^{2}\sigma,\label{mass relation}
\end{equation}
which can be used to eliminate the time-dependent mass $M_{-}$ out
of Eq. (\ref{effective_potential}). Upon direct substitution from
Eqs. (\ref{Vaidya_metric}) and (\ref{Schwarzschild function}) into
Eq. (\ref{effective_potential}) and then Eq. (\ref{mechanical_energy_equation}),
one arrives at 
\begin{equation}
\dot{a}=\sqrt{\frac{2M_{+}}{a}+4\pi a\sigma\left(\pi a\sigma-1\right)},\label{time evolution of the radius}
\end{equation}
for the time-evolution of the radius of the shell. 

\section{The Evolution of the Planetary Nebula}

Let us define for the energy density $\sigma$, a spherically symmetric
environment given by a simple power function of the form 
\begin{equation}
\sigma\equiv c_{0}a^{\mu},\label{energy density function}
\end{equation}
with $c_{0}$ being a positive constant and $\mu$ an integer. The
positivity of the constant $c_{0}$ is a must since we would like
the fluid on the shell to have a positive energy density and satisfies
the weak energy condition (WEC). In the case of a thin-shell, the
WEC states $S_{ij}V^{i}V^{j}\geq0$, in which $S_{ij}$ is the energy-momentum
tensor and $V^{i}$ is an arbitrary timelike vector. In the context
of perfect fluids, the WEC translates to two simultaneous conditions
$\sigma\geq0$ and $\sigma+p\geq0$. The satisfaction of the WEC guarantees
the ordinariness of the matter at the shell (Otherwise, the matter
will be the unwanted ``exotic matter''). Besides this, there is
subtle condition imposing an upper bound over the value of $c_{0}$.
It can be shown, by inserting (\ref{energy density function}) and
(\ref{time evolution of the radius 2}) into (\ref{energy_density}),
that the equation holds true only if $\left\vert c_{0}\right\vert \leq\left(2\pi a^{\mu+1}\right)^{-1}$.
Therefore, we have $\left(0,\left(2\pi a^{\mu+1}\right)^{-1}\right]$
as the permissible domain of $c_{0}$. However, note that the radius
of the nebular shell evolves by time, i.e. $a\equiv a\left(\tau\right)$.
As we now, a PN occurs in final stages of a highly evolved Sun-like
star and disappears in a few thousand years, leaving a white dwarf
behind. Let us assume that our model works to the moment $\tau_{f}$,
when the radius of the nebula is $a_{f}$ and it is not observable
in visible light anymore. This happens when the PNN cools down after
the fusion has almost stopped, so that it does not emit enough ultraviolet
radiation to ionize the distant nebular gas anymore. Accordingly,
$M_{-}$ approaches a final value, say $\left(M_{-}\right)_{f}$,
and the Vaidya metric becomes Schwarzschild. By this assumption, we
make sure that the conservation of mass and energy are satisfied.
In this limit, of course, the PN does not belong to the central mass
(the white dwarf) anymore and is part of the interstellar medium.
Hence, to make sure that $c_{0}$ remains bounded within its permissible
domain at all time during the expansion, we require that $c_{0}\in\left(0,\left(2\pi a_{f}^{\mu+1}\right)^{-1}\right]$.
Moreover, as a result of our considerations of the inner and outer
spacetimes, the mass of the central white dwarf $M_{-}$ decreases
by time, whereas the reduction is added to the mass of the nebula
through the outgoing waves, such that the total mass remains constant
($M_{+}$). Based on this, we should have $\mu\geq-2$ since for $\mu<-2$
the mass of the shell, i.e. $4\pi a^{2}\sigma$, decreases by time.
On the other hand, unlike $M_{+}$, $M_{-}$ is not a constant. The
central mass of the Vaidya metric $M_{-}$ (the mass of the PNN) constantly
gives off outgoing waves and shrinks, while its emitted energy is
absorbed by the shell and pushes it into the outer space. Hence, $\mu$
cannot be $-2$, as well, since it leads to a constant $M_{-}$, according
to Eq. (\ref{mass relation}). Therefore, we have $\mu>-2$ imposed
on the values of $\mu$. Note that, $\sigma$ is the energy density
of the shell not its mass density. So, although the mass density of
the shell decreases as the shell expands, the same is not necessarily
true for the energy density. In this seance, no upper bound is imposed
over the values of $\mu$.
\begin{figure}
\caption{\protect\includegraphics[scale=0.6]{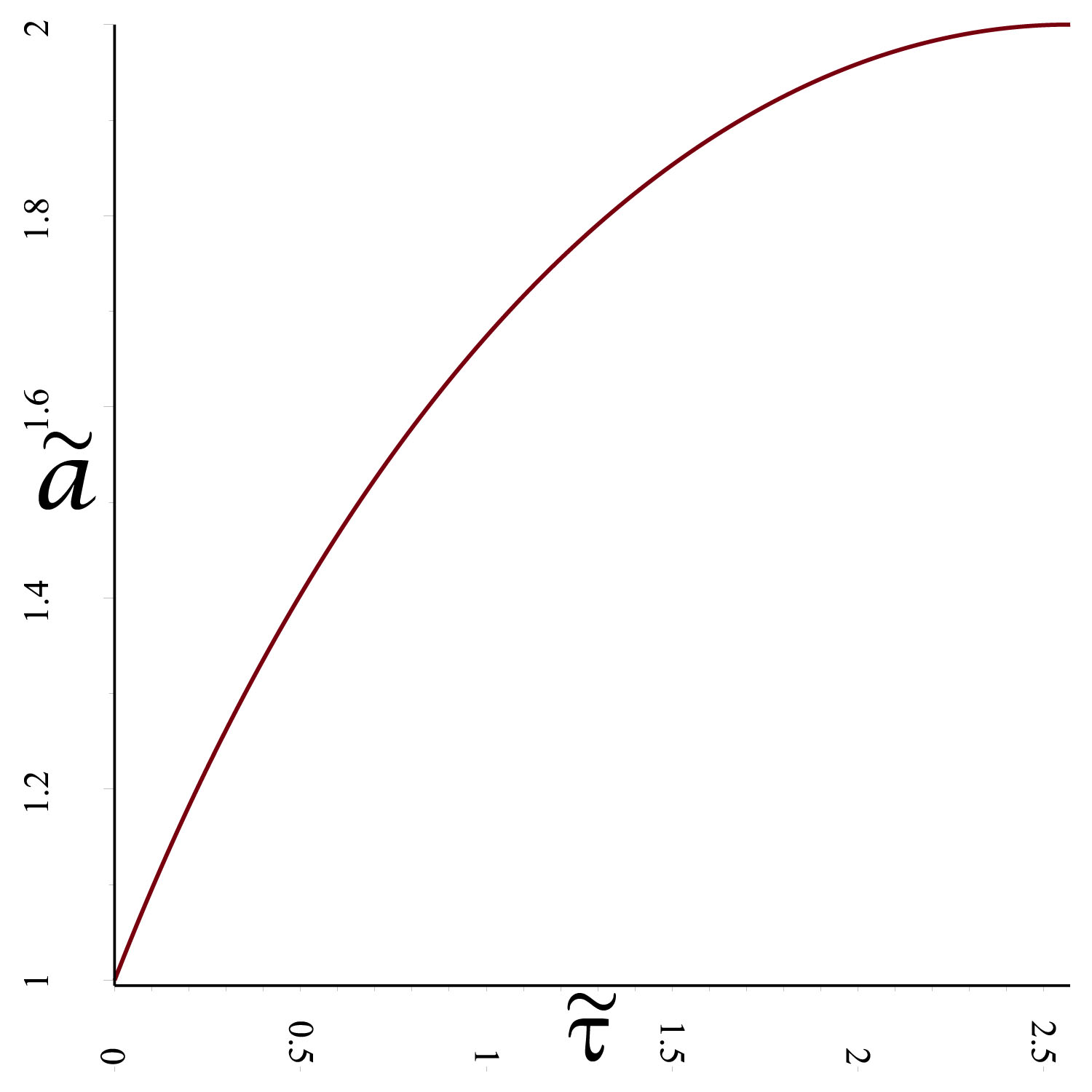}}
The plot of $\tilde{a}$ in terms of $\tilde{\tau}$ with $\tilde{a}_{0}=1$.
While the rescaled radius of PN increases its speed decreases and
when its speed becomes zero it instantly vanishes.
\end{figure}

As the first example, let us consider an ever decreasing energy density
by evoking $\mu=-1$. By looking at Eq. (\ref{time evolution of the radius}),
one sees that this choice has great mathematical advantages since
it simplifies Eq. (\ref{time evolution of the radius}) to 
\begin{equation}
\dot{a}=\sqrt{\frac{2M_{+}}{a}-A},\label{time evolution of the radius 2}
\end{equation}
where $A\equiv4\pi c_{0}\left(1-\pi c_{0}\right)$ is just a constant.
Hereon, we will refer to $A$ as the evolution constant. According
to the limitation over $c_{0}$, for the evolution parameter we always
have $0<A\leq1$ (considering the positivity of $c_{0}$). This differential
equation is analytically solvable, although it cannot be written explicitly
for $a\left(\tau\right)$. The answer is the solution of the algebraic
equation 
\begin{equation}
\tau+\frac{\sqrt{Aa\left(\tau\right)\left(2M_{+}-Aa\left(\tau\right)\right)}-M_{+}\sin^{-1}\left(\frac{Aa\left(\tau\right)}{M_{+}}-1\right)}{A^{3/2}}+C=0\label{general solution}
\end{equation}
where $C$ is an integral constant that can be determined by the initial
condition $a\left(0\right)=a_{0}$. Here, $a_{0}$ is the radius of
the asymptotic giant branch (AGB) star when its outermost layers eject
and form the expanding envelope of the PN. According to the model,
although the mass $M_{+}$ and the evolution constant $A$ are different
for different PNe, almost the same evolution pattern is expected.
This, of course, can be seen by imposing the rescaling 
\begin{equation}
\tilde{a}\equiv\frac{Aa}{M_{+}}\qquad\text{and}\qquad\tilde{\tau}+\tilde{C}\equiv\frac{A^{\frac{3}{2}}}{M_{+}}\text{\ensuremath{\left(\tau+C\right)},}\label{rescaling}
\end{equation}
which cast the implicit equation in (\ref{general solution}) into
\begin{equation}
\tilde{\tau}+\tilde{C}=\sin^{-1}\left(\tilde{a}-1\right)-\tilde{a}\sqrt{\frac{2}{\tilde{a}}-1}.\label{rescaled general solution}
\end{equation}
We add that, $\tilde{C}$ is obtained using the initial condition
i.e., $\tilde{a}\left(0\right)=\tilde{a}_{0}$ upon which 
\[
\tilde{C}=\sin^{-1}\left(\tilde{a_{0}}-1\right)-\tilde{a_{0}}\sqrt{\frac{2}{\tilde{a_{0}}}-1}.
\]
This new equation is free of parameters $M_{+}$ and $A$, and reflects
the general behavior of the evolution of the radius of spherical PNe.
In Fig. 1, we plot the rescaled radius $\tilde{a}$ against $\tilde{\tau}$.
For a particular nebula, with a specific evolution parameter $A$,
we must have $\tilde{a}>2A$ at all times, since $\tilde{a}=2A$ corresponds
to the Schwarzschild radius of the exterior spacetime. 

\section{Conclusion}

We established a geometrical model for the evolution of PN. After
deriving the energy density and pressure expressions for the thin-shell
we obtained the time evolution of the thin-shell's radius. We plotted
the rescaled radius as a function of rescaled time for an energy density
obeys a simple power function. While the radius of thin-shell increases
its speed decreases. In its final stage the speed of PN becomes zero
and it disappears instantly. Although, the model presented seems analytically
reasonable it needs observational verification as well as to see a
simulation of the model might give some clues about the evolution
of PN.

\end{document}